\documentclass[aps,prb,10pt,twocolumn,superscriptaddress]{revtex4-1}
\usepackage{graphicx}
\usepackage{amsmath,amssymb,bm}
\usepackage{color}
\usepackage[colorlinks=true,citecolor=blue,linkcolor=red]{hyperref}

\newcommand{\Ec}{E_\mathrm{c}}
\newcommand{\Ur}{U_\mathrm{r}}
\newcommand{\rhoC}{\rho_\mathrm{c}}
\newcommand{\rhot}{{\tilde \rho}}
\newcommand{\bmk}{{\bm k}}
\newcommand{\kB}{k_\mathrm{B}}
\begin{document}

%==============================================================================================
\title{
Electrical permittivity driven metal-insulator transition in heterostructures of nonpolar Mott- and band insulators
}

\author{Yukiko Omori}
\affiliation{Institute for Advanced Research, Nagoya University, Nagoya 464-8602, Japan}
\affiliation{Department of General Education, Toyota National College of Technology, Toyota 471-8525, Japan}

\author{Andreas R\"uegg}
\affiliation{Theoretische Physik, ETH Z\"urich, CH-8093 Z\"urich, Switzerland}

\author{Manfred Sigrist}
\affiliation{Theoretische Physik, ETH Z\"urich, CH-8093 Z\"urich, Switzerland}

\date{\today}

\begin{abstract}
Metallic interfaces between insulating perovskites are often observed in heterostructures combining polar and nonpolar materials. In these systems, the polar discontinuity across the interface may drive an electronic reconstruction inducing free carriers at the interface. Here, we theoretically show that a metallic interface between a Mott- and a band-insulator can also form in the {\em absence} of a polar discontinuity. The condition for the appearance of such a metallic state is consistent with the classical Mott criterion: the metallic state is stable if the screening length falls below the effective Bohr radius of a particle-hole pair. In this case, the metallic state bears a remarkable similarity to the one found in polar/nonpolar heterostructures. On the other hand, if the screening length approaches the size of the effective Bohr radius, particles and holes are bound to each other resulting in an overall insulating phase. We analyze this metal-insulator transition, which is tunable by the dielectric constant, in the framework of the slave-boson mean-field theory for a lattice model with both onsite and long-range Coulomb interactions. We discuss ground-state properties and transport coefficients, which we derive in the relaxation-time approximation. Interestingly, we find that the metal-insulator transition is accompanied by a strong enhancement of the Seebeck coefficient in the band-insulator region in the vicinity of the interface. The implications of our theoretical findings for various experimental systems such as nonpolar (110) interfaces are also discussed.
\end{abstract}

\maketitle
%==============================================================================================

%=================================================================
\section{introduction}
%=================================================================

Recent advances in the crystal growth techniques allowed researchers to fabricate precise
heterojunctions of different transition metal oxides.\cite{Chakhalian:2012,Hwang:2012}
Heterostructures and superlattices exhibit various properties, which dramatically differ from those of a single bulk substance,
yielding new possibilities to study electron correlations and potential novel devices.
One of the most renowned phenomena is the emergence of a metallic quasi-two-dimensional electron gas 
at the interface between two different insulators.
Since Ohtomo and co-workers found metallic properties in
LaTiO$_3$ (LTO)/SrTiO$_3$ (STO) (Ref.~\onlinecite{Ohtomo2002}) and 
LaAlO$_3$ (LAO)/STO (Ref.~\onlinecite{Ohtomo2004}), the occurrence of conducting interfaces with high carrier numbers has frequently been reported
for those heterostructures \cite{Shibuya2004,Takizawa2006,Seo2007,Thiel2006,Joshua:2012,Berner:2013,Annadi:2013} as well as for other material combinations \cite{Boris:2011,Zhang:2013,Liu:2013}.
As explanations for the formation of the metallic interface, several possibilities have been proposed, including  
carrier doping from oxygen vacancies \cite{Ohtomo2004,Pentcheva2006,Park2006,Thiel2006,Takizawa2006},
lattice relaxations \cite{Okamoto2006,Hamann2006,Ishibashi2008,Ishibashi2010,Maurice2006,Willmott2007},
and interfacial roughening \cite{Willmott2007,Kourkoutis2007}.

In this context, an important observation was that the metallic state usually forms at interfaces between a
polar and a nonpolar insulator of the ABO$_3$ perovskite-type.\cite{Ohtomo2004,Nakagawa2006,Hwang2006,Pauli2008, Moetakef:2011,Moetakef:2012,Bristowe:2014} A first understanding of this observation is gained by considering the charge distribution across such a heterostructure using an ionic picture with doubly charged negative oxygen ions O$^{2-}$. In this picture, heterostructures grown along the cubic (001) direction are viewed as a stack of alternating AO and BO$_2$ layers.
Materials such as STO are of the type A$^{2+}$B$^{4+}$O$_3$ and therefore only contain neutral atomic layers, i.e.,\, these are {\em nonpolar} materials.
On the other hand, materials of the type A$^{3+}$B$^{3+}$O$_3$ are {\em polar} along the (001) direction in the sense that they 
consist of alternating charged planes (AO)$^{+}$ and (BO$_2$)$^{-}$. LTO and LAO belong to this class of materials. The polar and nonpolar situation described above also correspond to the two values of the formal bulk polarization allowed by the cubic symmetry, which for band insulators can be directly obtained via the calculation of the Berry phase.\cite{Murray:2009,Bristowe:2014} From the ionic picture we now see that a polar discontinuity arises at the interface in the LTO/STO and LAO/STO heterostructures. Such a situation would cause the so-called polar catastrophe, if the width of the polar material is increased. To avoid this huge energy penalty, the systems reacts by charging the top surface and interface by an amount of $\pm e/2$ per surface unit cell. For the top surface, this usually happens via atomic reconstruction, resulting in a charged but insulating surface. For the interface, however, it was suggested \cite{Ohtomo2004,Nakagawa2006,Hwang2006,Pauli2008} that a more interesting situation can occur, which is called electronic reconstruction. In this scenario, the additional electronic charge $e/2$ resides on the transition-metal ions near the interface, leading to a mixed valence state (e.g.\ Ti$^{4+}$/Ti$^{3+}$) with metallic properties.

While the scenario of the electronic reconstruction driven by the polar discontinuity provides a robust criterion to identify interfaces with expected metallic properties, it can not make predictions about interfaces which lack a polar discontinuity. 
Clearly, in the ionic picture, such interfaces are expected to be insulating. 
However, unexpected interfacial conductivity was observed recently in the LAO/STO heterostructure with a (110) growth direction.\cite{Annadi:2013}
This is interesting because the ideal (110) system does not have a polar discontinuity. Instead, the ionic picture predicts a polar/polar situation with alternating $\mathrm{(ABO)}^{4+}$ and $\mathrm{(O_2)}^{4-}$ planes for any A and B atoms. Nevertheless, it was found that the (110) interface bears a great resemblance to the (001) structure in the conductivity, carrier density, Hall mobility, and even the critical width for the metal-insulator transition.\cite{Annadi:2013} One possible reason for this resemblance is that the (110) interface is non-ideal and formed by several plateaus of the (001)-type interface.\cite{Annadi:2013} But there is also the possibility that the ionic picture fails and that a metallic state is formed at the ideal interface in the absence of a polar discontinuity as a result of the covalent character of the transition-metal oxides. It is intuitively clear that enhanced charge fluctuations near the interface, which result in a smooth electronic charge distribution, can lead to metallicity, irrespective of the polar or non-polar nature of the involved materials.

The goal of this paper is to explore this latter possibility, namely the formation of a metallic state in heterostructures {\em without} polar discontinuity. In particular, we focus on the role of the dielectric constant, which is an important parameter in determining the smoothness of the global charge distribution. For this purpose, we theoretically investigate 
a lattice model of a Mott-insulator (MI)/ band-insulator (BI) heterostructure, which consists of only 
electrically neutral layers along the growth direction, 
i.e., the {\em nonpolar/nonpolar} structure,
and contrast these results to earlier studies of polar/nonpolar heterostructures.\cite{Okamoto2004_nature428,Okamoto2004_PRB70,Okamoto2004_PRB70R,Okamoto2005,Okamoto2006,Kancharla2006,Lee2006,Lee2007,
Ruegg2007,Ruegg2008,Ishida2008,Ishida2009,Ishida2010,Ueda2012,Lechermann:2013} We examine the ground-state and transport properties in the 
absence of the polar discontinuity, using the Kotliar-Ruckenstein slave-boson mean-field (SBMF) theory
\cite{Kotliar1986,Ruegg2007,Ruegg2008} to treat the onsite and long-range Coulomb interactions. We identify two different regimes in accordance with Mott's classical criterion: \cite{[{See, for instance,  }]DuanGuojun:2005} first, if the screening length is smaller than the effective Bohr radius, both the polar/nonpolar and the nonpolar/nonpolar MI/BI heterostructures exhibit an interfacial metallic state. In this case, the transport properties are rather similar for the two cases. Second, if the screening length exceeds the effective Bohr radius, the nonpolar heterostructures undergo a metal-insulator transition while the polar/non-polar interfaces remain metallic. The metal-insulator transition in the nonpolar system is tunable by the electrical permittivity, which controls the ``sharpness'' of the electronic charge distribution across the interface. In some materials, such as STO, the static dielectric constant strongly varies with temperature and electric field, potentially allowing to tune this transition externally. Interestingly, we find that 
the metal-insulator transition is accompanied by a strong enhancement of the thermoelectric response in the vicinity of the interface.

This paper is organized as follows.
In Sec.~\ref{sec:main_results} we summarize the main results of our paper.
In Sec.~\ref{sec:model} we give a detailed discussion of the model and used method.
Sections \ref{sec:ground_state} and \ref{sec:MI_transition} are devoted to show detailed results:
The SBMF ground state properties are presented in the former section, and the transport properties are discussed in the latter.
We conclude in the final Sec.~\ref{sec:conclusion}.

%=================================================================
\section{Main results}
\label{sec:main_results}
%=================================================================
Before we discuss our main results, we briefly describe the considered model. The mathematical details are provided in Sec.~\ref{sec:model}. 
\begin{figure}
 \centering
 \includegraphics[width=0.45\textwidth]{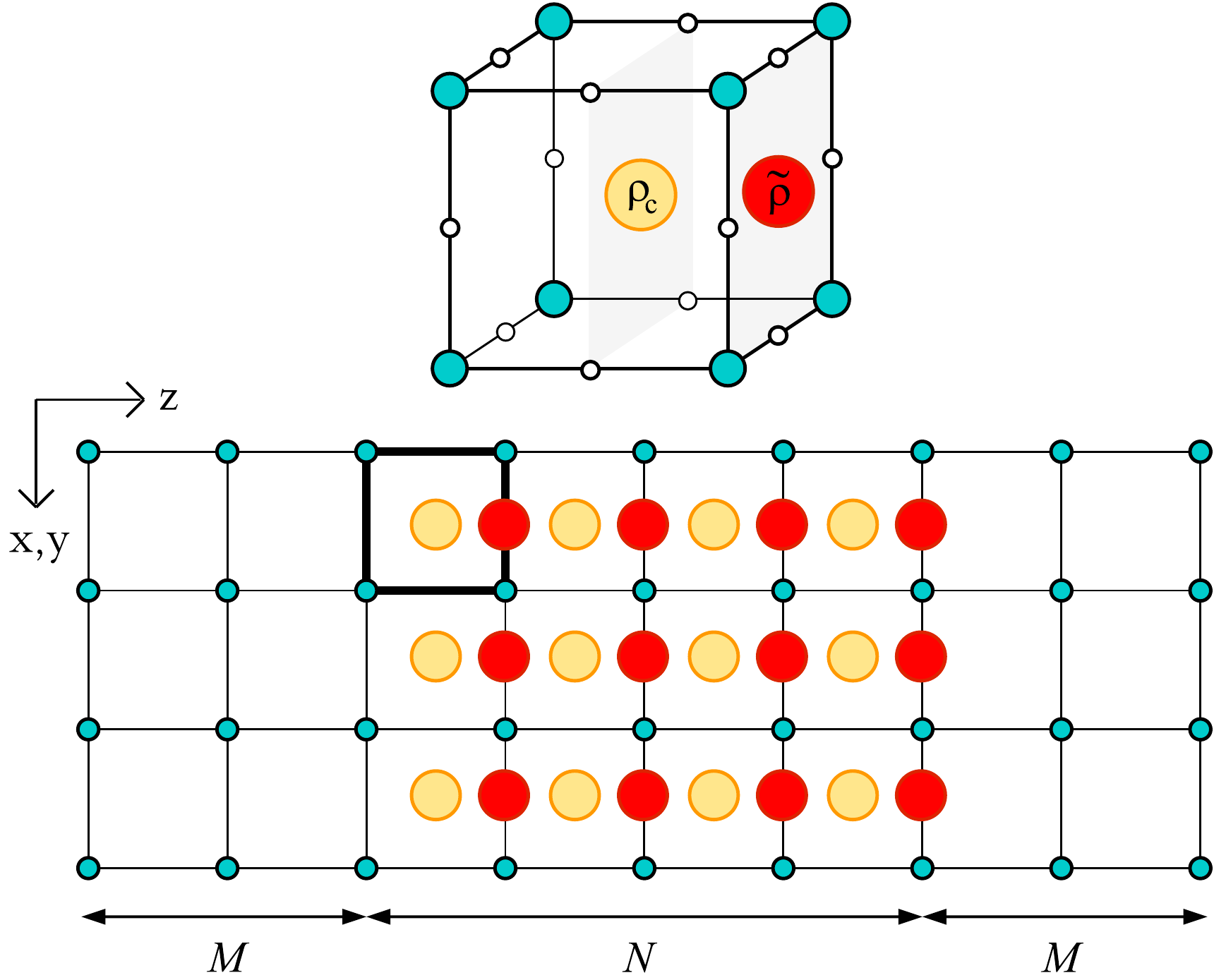}
 \caption{
 (Color online.)
 Schematic view of the heterostructure studied in this paper.
 The upper figure is the three-dimensional view of a unit block indicated by the thick lines in the lower.
 The blue circles represent the electron sites.
 }
 \label{fig:model}
\end{figure}
We study a lattice model of a BI/MI/BI sandwich structure. Figure~\ref{fig:model} schematically shows the electronic sites along the z axis with $N$ layers of positive point-like background charges $\rho_{\mathrm c}$ and $\rhot$, which are located in-between the electronic sites and satisfy $\rho_\mathrm{c} + \rhot = 1 $ in units of $e$. We consider charge neutral systems; thus, the uniform MI material has precisely one electron per site while the uniform BI is modeled by an empty conduction band. The model describes a polar/nonpolar heterostructure if the positive background charges are located in the center of the cubes formed by the electronic sites: $\rho_c=1$ and $\rhot=0$. In this case, the MI is polar along the growth direction with alternating positively and negatively charged layers. On the other hand, the model describes a nonpolar/nonpolar heterostructure if we shift the positive charges into the electronic layers: $\rho_c=0$ and $\rhot=1$, cf.~Fig.~\ref{fig:model}. In this case, the MI is non-polar with charge neutral layers along the growth direction. Besides the nearest-neighbor hopping amplitude $t$, two more electronic energy scales are introduced: (i) a local electron-electron repulsion $U$ of the Hubbard type and (ii) an energy scale $\Ec=e^2/(\varepsilon a)$ characterizing the long-range Coulomb interaction, where $\varepsilon$ is the dielectric constant and $a$ the lattice constant. 

\begin{figure}
 \centering
 \includegraphics[]{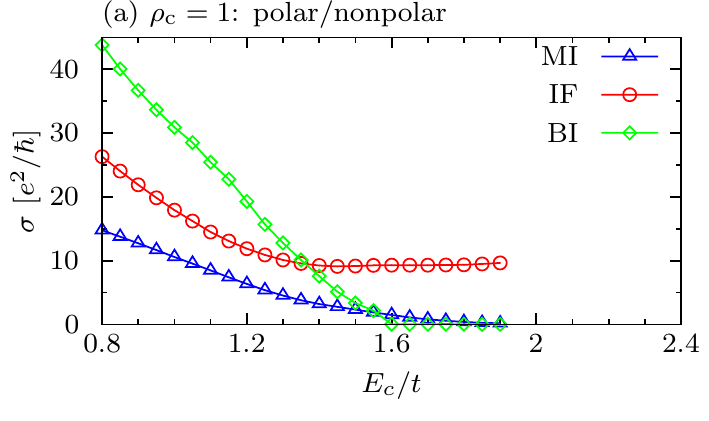}
 \includegraphics[]{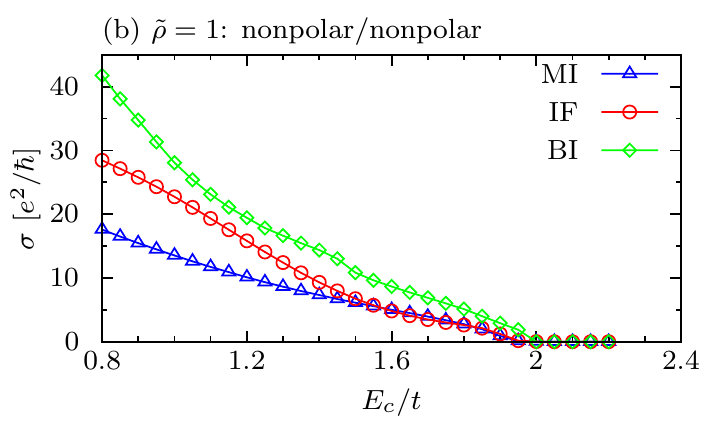}
 \caption{
 (Color online.)
 The contributions from the MI, the interface (IF), and the BI region to the total conductivity in the large-$U$ limit ($\Ur=U-\Ec=25t$) as function of the parameter $\Ec$ for 
 (a) the polar/nonpolar and (b) the nonpolar/nonpolar heterostructures with $N=10$. The presented IF conductivity is a sum over three (two) layers around the interface for $\rhoC=1$ ($\rhot=1$), and the MI and BI conductivities are obtained from sums over the remaining inside and outside layers, respectively.}
 \label{fig:sigma_rhot0-1}
\end{figure}
One important physical difference between the polar/nonpolar ($\rhoC=1$) and the nonpolar/nonpolar ($\rhot=1$) heterostructure in the the large $U$ limit becomes apparent if we consider the spatially resolved electric conductivity in the direction perpendicular to the growth direction as function of $\Ec$, see Fig.~\ref{fig:sigma_rhot0-1}. In the small $\Ec$ limit, electrons from the MI region substantially leak into the BI region resulting in a relatively smooth charge distribution. As a result, not only the interface layers but also regions in the BI and MI significantly contribute to the transport. Remarkably, the spatially resolved conductivity of the $\rhot=1$ nonpolar heterostructure shows a high resemblance to the one of the $\rhoC=1$ polar/nonpolar structure for small $\Ec$. The charge distribution becomes sharper for increasing $\Ec$ and the spatially resolved conductivities start to differ between the two heterostructures. In the $\rhoC=1$ polar/nonpolar heterostructure, the MI and BI conductivities approach zero around $\Ec=1.6t$ due to the full occupation of the lower Hubbard band in the MI region and the loss of the itinerant electrons in the BI region. However, the interface remains metallic with the electronic charge density $n \sim 0.5$ per surface area required by charge neutrality. This interfacial charge is a consequence of the polar/nonpolar nature of the heterostructure and leads to transport dominated by the interface layers. By contrast, in the $\rhot=1$ nonpolar/nonpolar heterostructure, the interfacial metallicity gradually disappear with increasing $\Ec$, and the whole system undergoes a metal-insulator transition at $\Ec=1.96t$.

\begin{figure}
 \centering
 \includegraphics[]{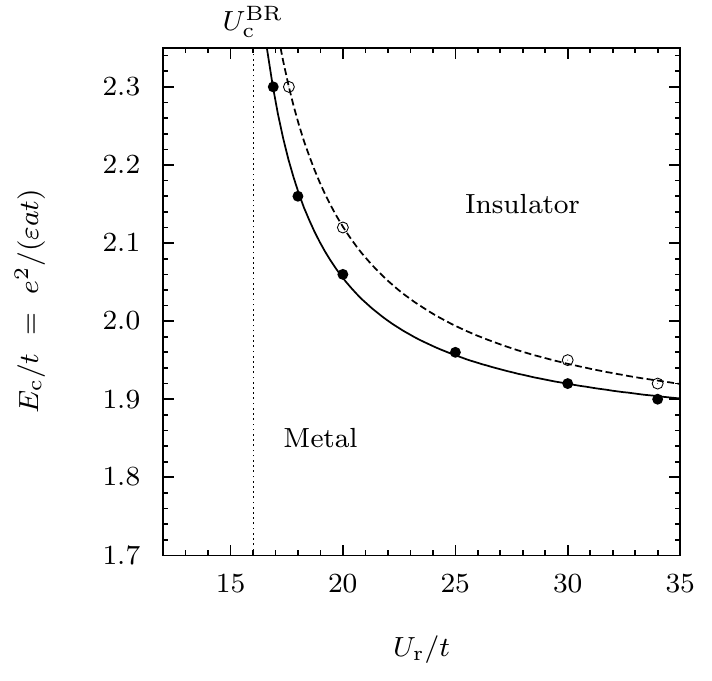}
 \caption{
 The $\Ur$-$\Ec$ phase diagram for the $\rhot=1$ nonpolar/nonpolar heterostructure showing the metallic and insulating phases.
 The solid and dashed lines denote the boundaries between metallic and insulating states for $N=10$ and $N=2$ numbers of MI layers, respectively.
 }
 \label{fig:PD}
\end{figure}
This transition is specific to the nonpolar/nonpolar heterostructure with exactly $\rhot=1$ and depends both on the value of $U$ and $\Ec$.
Figure \ref{fig:PD} shows the metal-insulator phase diagram in the $\Ur$-$\Ec$ plane ($\Ur=U-\Ec$) for two different system sizes. The insulating regime requires both a large $\Ur$ and a large $\Ec$. A qualitative understanding of the phase diagram can be obtained by considering the two relevant length scales in the problem. First, in the metallic phase, the Coulomb interaction is screened over a scale estimated from the Thomas-Fermi length,
\begin{equation}
 \lambda_\mathrm{TF}
  = \sqrt{ \frac{\varepsilon}{4\pi e^2} \frac{\partial \mu_0}{\partial n}}
  = \sqrt{ \frac{1}{4\pi a \Ec {\bar \kappa}} },
  \label{eq:TF}
\end{equation}
where $\mu_0$ and $n$ are the chemical potential and the density of the free electrons in the homogeneous system, respectively,
and ${\bar \kappa} = \partial n / \partial \mu_0$. Second, the effective Bohr radius of an exciton
\begin{equation}
a_\mathrm{B} = \frac{\varepsilon \hbar^2}{m^* e^2} \sim \frac{2mt}{m^*\Ec} a,
\label{eq:a_B}
\end{equation}
describes the extension of a bound particle-hole pair. Mott now argued that a metallic state requires $\lambda_\mathrm{TF}<a_\mathrm{B}$ \cite{DuanGuojun:2005}, or, in other words, the Coulomb potential has to be sufficiently weak in order not to bind a particle and a hole. Applying Mott's criterion with the above estimates for the characteristic length scales results in the condition $\Ec/t \lesssim 16 \pi t a^3 (m/m^*)^2 {\bar \kappa} $ for metallic behavior. The function $\bar \kappa$ and the inverse of the effective mass $m/m^*$ decrease with increasing $U$, and consequently the critical $\Ec$ becomes a decreasing function of $U$ as observed in Fig.~\ref{fig:PD}.

\begin{figure}
 \centering
 \includegraphics[]{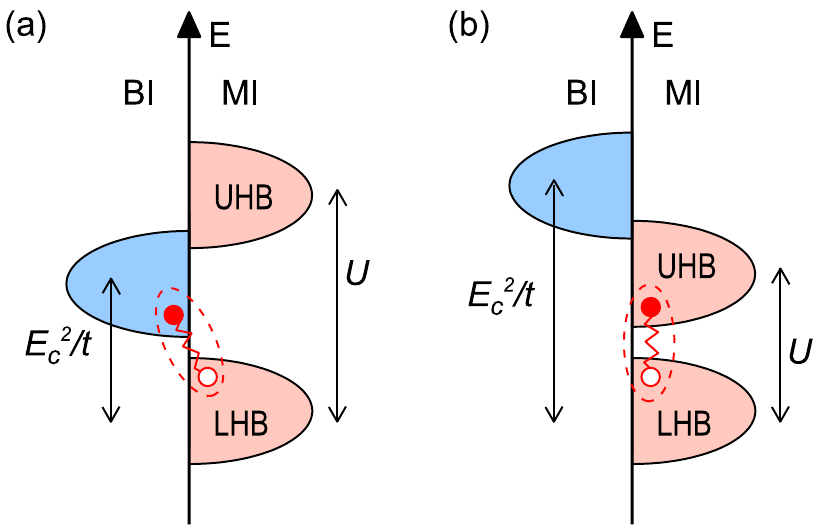}
 \caption{
 Schematic energy diagram for the insulator with (a) large $\Ec$ and small $U$, and (b) large $U$ and small $\Ec$.
 }
 \label{fig:DOS}
\end{figure}
\begin{figure}
 \centering
 \includegraphics[]{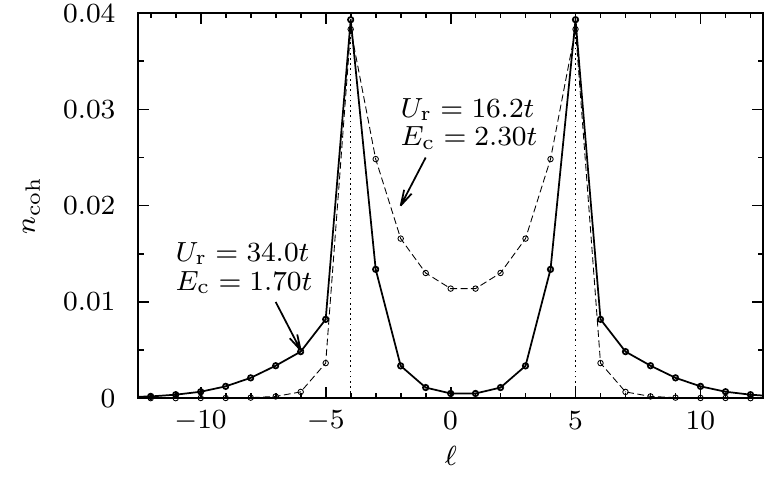}
 \caption{
 Ditributions of the coherent particle densities $n_\mathrm{coh}$.
 The solid and dashed lines indicates $(\Ur,\Ec)=(34.0,1.70)t$ and $(16.2,2.30)t$, respectively.
 }
 \label{fig:Ncoh}
\end{figure}

We can gain further understanding by considering the insulating phase in the two extreme limits (i) $\Ur\rightarrow \infty$ and (ii) $\Ec\rightarrow\infty$. In the first regime, where the local Coulomb repulsion $U$ dominates over the kinetic energy $W$, particle-hole excitations predominantly occur between the BI and the lower Hubbard band of the MI material, as illustrated in Fig.~\ref{fig:DOS}(a). Consequently, the phase boundary in this regime is roughly independent of $U$ and the transition is driven by the electric permittivity, alone. Furthermore, on the metallic side of the transition, we expect that the free carriers are predominantly located in the BI region close to the interface. This is indeed observed in the layer-resolved coherent charge density $n_\mathrm{coh}$ shown in Figure \ref{fig:Ncoh} [the definition of $n_{\rm coh}$ is given in Eq.~\eqref{eq:ncoh}].
On the other hand, in the parameter regime with large $\Ec$, the transition is essentially governed by the suppression of the local charge fluctuations in the central region of the heterostructure. It corresponds to the usual Mott insulator situation where particles are excited across the Mott-Hubbard gap [Fig.~\ref{fig:DOS}(b)]. 
Hence, the critical $\Ur$ becomes independent of $\Ec$ and saturates roughly at the metal-insulator transition of the bulk system (in our approximation given by the Brinkman-Rice transition point \cite{Brinkman:1970}, $U_\mathrm{c}^\mathrm{BR} \sim 16t$). Furthermore, the coherent particles on the metallic side of the transition are predominantly located in the MI region, as can be seen in Fig.~\ref{fig:Ncoh}.

\begin{figure}
 \centering
 \includegraphics[]{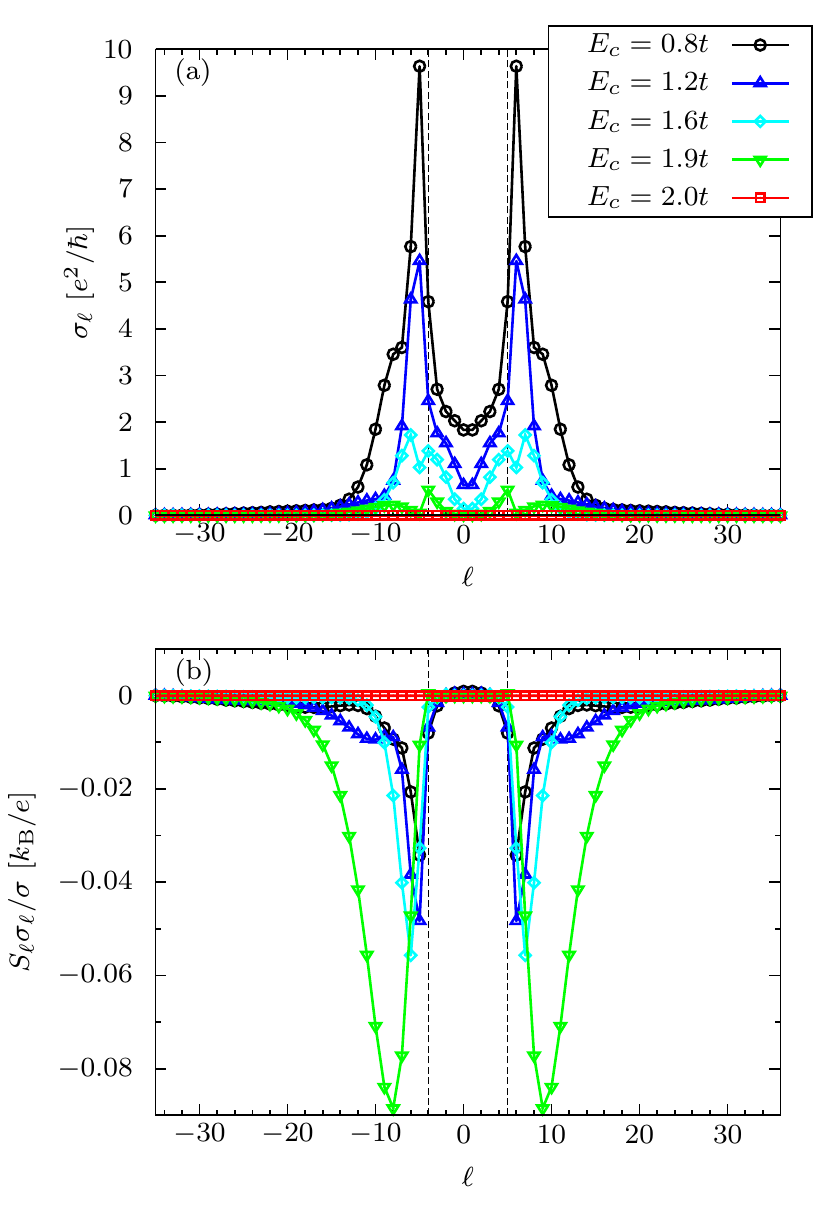}
 \caption{
 (Color online)
 The distributions of the layer-resolved conductivity $\sigma_\ell$ and the contribution to the total Seebeck coefficient $S_\ell \sigma_\ell/\sigma$
 in the $\rhot=1$ nonpolar/nonpolar heterostructure
 with $\Ur=25t$ and $\Ec=0.8t,\cdots,2.0t$.
 The dashed lines indicate the position of the interfaces, $\ell=-4$ and $5$.
 }
 \label{fig:LayerTransports}
\end{figure}

The electric permittivity driven metal-insulator transition is accompanied by characteristic features in the layer-resolved electrical and thermal transport. 
We illustrate this aspect in Fig.~\ref{fig:LayerTransports}, where the layer-resolved contributions to the total conductivity $\sigma=\sum_\ell\sigma_\ell$ and total Seebeck coefficient $S= \sum_\ell S_\ell \sigma_\ell/\sigma$ are presented for various $\Ec$ with $\Ur=25t$.
For small $\Ec$, the main contribution to $\sigma$ comes from the layers in the BI region neighboring the interface.
With increasing $\Ec$, the electrical conductivity is suppressed and tends to zero if the metal-insulator transition at $\Ec=1.96t$ is approached.
The Seebeck coefficient also has a maximum on the same layers for $\Ec=0.8t$ [Fig.~\ref{fig:LayerTransports}(b)]. However, in contrast to $\sigma_\ell$, the peak in $S_\ell\sigma_\ell/\sigma$ gradually moves out into the BI region and experiences a strong enhancement toward the metal-insulator transition.
This behavior is caused by the reduction of free carriers in the BI region: near the transition point, the chemical potential approaches the band edge of the bands describing the weakly bound states in the BI region. This causes the reduction of the conductivity and the enhancement of the Seebeck coefficient. The spatial distribution of $S_\ell\sigma_\ell/\sigma$ essentially follows the shape of the wave function, as discussed in more details in Sec.~\ref{sec:loc_and_ren}.

In the following sections, we present details of our models, methods, and numerical results
in the $\rhot=1$ nonpolar heterostructure and clarify the origin of the metallic interface.

%=================================================================
\section{Model and formalism}
\label{sec:model}
%=================================================================

%-----------------------------------------------------------------
\subsection{Model of a nonpolar BI/MI/BI heterostructure}
%-----------------------------------------------------------------

Our model for the BI/MI/BI heterostructure is given by a generalized Hubbard model
\begin{equation}
 H = 
  -t \sum_{\left<i,j\right>,\sigma} \hat{c}_{i,\sigma}^\dagger \hat{c}_{j,\sigma}
  + U \sum_i \hat{n}_{i,\uparrow} \hat{n}_{i,\downarrow}
  + \sum_i \phi_i \hat{n}_i ,
  \label{eq:H}
\end{equation}
where $\hat{c}_{i\sigma}^{(\dag)}$ is the creation (annihilation) operator for an electron with spin $\sigma$ on site $i$, $\hat{n}_{i\sigma}=\hat{c}_{i\sigma}^{\dag}\hat{c}_{i\sigma}$ and $\hat{n}_i=\hat{n}_{i\uparrow}+\hat{n}_{i\downarrow}$. In transition-metal oxide heterostructures, the relevant orbital degrees of freedom are usually the $d$-orbitals. In model \eqref{eq:H}, we disregard this additional complication and consider a single-orbital model, instead. Although the orbital physics is a crucial ingredient to describe real oxide heterostructures, it is not important for addressing the differences or similarities between nonpolar and polar heterostructures, which is the main focus of this paper. 

The first term of model \eqref{eq:H} describes the kinetic energy from nearest-neighbor hopping processes with transfer energy $t$. The second term describes the onsite repulsion with Hubbard interaction $U$. Finally, the third term describes the long-range Coulomb interaction: The quantity $\phi_i$ is the electrostatic potential at the electron site ${\bm R}_i = a{\bm r}_i$,
and obeys Poisson's equation. Its discretized solution for the point-like charges in our system is
\begin{equation}
 \phi_i
  =
    \Ec \sum_{j(\neq i)} \frac{n_j}{|{\bm r}_i - {\bm r}_j|}  
  - \Ec \sum_{j}^\mathrm{ion} \frac{\rho_{j}}{|{\bm r}_i - {\bm r}_j^\mathrm{ion}|} .
  \label{eq:phi}
\end{equation}
For simplicity, we assume a constant $\Ec = e^2/\varepsilon a$ over the whole lattice. The first term in Eq.~(\ref{eq:phi}) is the electron-electron Coulomb interaction and $n_j=\left< \hat{n}_j \right>$.
The second is the electron-ion attraction with the positively-charged ions $\rho_j$ at ${\bm R}_j^\mathrm{ion} = a{\bm r}_j^\mathrm{ion}$.
As indicated in Fig.~\ref{fig:model} and described in Sec.~\ref{sec:main_results}, we take $\rho_j=\rhoC$ or $\rhot$ according to the position ${\bm r}_j^\mathrm{ion}$, and impose $\rhoC + \rhot = 1$. For the nonpolar heterostructure, $\rhoC=0$ and only the charges $\rhot$, which are located in the same $xy$ planes as the electron sites, are considered. On the other hand, the model with $\rhoC=1$ describes the polar/nonpolar heterostructure and was previously studied as a model of the LTO/STO heterostructure.\cite{Okamoto2004_PRB70,Ruegg2007}

%-----------------------------------------------------------------
\subsection{Slave-boson mean-field treatment}
%-----------------------------------------------------------------

We treat the model Hamiltonian (\ref{eq:H}) 
within the paramagnetic Kotliar-Ruckenstein SBMF method following the formalism introduced in Ref.~\onlinecite{Ruegg2007} for the polar/nonpolar heterostructure, which also provides additional details. In the slave-boson treatment, the Fock space is enlarged to contain a set of two fermions and four bosons at each site
with the pseudofermionic operator $\hat{f}_{i\sigma}^{(\dagger)}$ and
the bosonic annihilation (creation) operators
$\hat{e}_i^{(\dagger)}$,  $\hat{p}_{i\sigma}^{(\dagger)}$, and $\hat{d}_i^{(\dagger)}$, 
which represent an empty, a singly occupied site with spin $\sigma$, and a doubly occupied site, respectively.
To eliminate unphysical states, constraints are imposed to these operators as described in Ref.~\onlinecite{Kotliar1986}.
The electron annihilation (creation) operator is mapped on the enlarged Fock space as
$ \hat{c}_{i\sigma}^{(\dagger)} \rightarrow (\hat{z}_{i\sigma} \hat{f}_{i\sigma})^{(\dagger)}$,
where the Kotliar-Ruckenstein bosonic operator $\hat{z}_{i\sigma}^{(\dagger)}$ 
ensures that the movement of the bosons follows the fermions.
Using this slave-boson representation,
we can reformulate the Hamiltonian (\ref{eq:H}) as
\begin{align}
 H^\mathrm{sb} =
 & -t \sum_{\left<i,j\right>,\sigma} 
 \hat{z}_{i\sigma}^\dagger \hat{z}_{j\sigma} \hat{f}_{i,\sigma}^\dagger \hat{f}_{j,\sigma}
 \nonumber \\
 & + \Ur \sum_i \hat{d}_{i}^\dagger \hat{d}_{i} 
   +  \sum_{i,\sigma}  \left( \phi_i + \frac{U_0}{2} \hat{n}_{i,\tilde{\sigma}}  \right) \hat{n}_{i,\sigma} ,
 \label{eq:Hsb}
\end{align}
where we redefine 
$ \hat{n}_{i,\sigma} = \hat{f}_{i,\sigma}^\dagger \hat{f}_{i,\sigma} 
= \hat{p}_{i\sigma}^{\dagger} \hat{p}_{i\sigma} + \hat{d}_i^\dagger \hat{d}_i $,
and $\tilde{\sigma} = \uparrow (\downarrow)$ for $\sigma=\downarrow(\uparrow)$.
Note that $\Ur=U-U_0$, and as shown below,
$U_0$ is the amount of the onsite repulsion treated within the Hartree approximation.
Although there is an ambiguity to settle the value of $U_0$,
we define $U_0=\Ec$ since the physical relevant parameter regime is expected to be $U \ge \Ec$.\cite{Ruegg2007}

In the mean-field approximation, the slave-boson fields are replaced by their mean values.
In order to investigate the essential aspects of the interfacial metallicity,
it is enough to concentrate on the paramagnetic solutions.
In addition, for solutions satisfying the lattice symmetry, all the mean fields depend only on the layer index $\ell$.
In this case, only three kinds of layer-dependent mean fields remain to be determined:
the electronic charge density $n_\ell$,
the amplitude of doubly occupancy $d_\ell$, and a Lagrange multiplier $\lambda_\ell$.
These fields are determined from the saddle point of the free energy,
\begin{align}
 f[n,d,\lambda] =
 &- \frac{2}{\beta N_\parallel} \sum_{\bmk,\nu} \ln \left( 1+ e^{-\beta E_{\bmk\nu}} \right)
 \nonumber \\
 &+ \Ur \sum_\ell d_\ell^2 - \sum_\ell \left( \lambda_\ell + \mu \right) n_\ell ,
\end{align}
where $\beta = 1/\kB T$ and $N_\parallel$ denotes the number of sites in a layer.
The chemical potential $\mu$ is determined by the condition of charge neutrality,
$\sum_\ell n_\ell = N$.
The quasiparticle energy spectrum $E_{\bmk\nu}$ with the in-plane wave vector $\bmk$ and 
a band index $\nu$ are obtained from an effective one-dimensional Schr\"odinger equation,
\begin{align}
 & E_{\bmk\nu} \psi_{\bmk\nu}(\ell)
 \nonumber \\
 =& \left\{ z_\ell^2 \epsilon_{\bmk} + \bar{\phi}_\ell + \lambda_\ell \right\} \psi_{\bmk\nu}(\ell)
 -t \sum_{\ell'=\pm 1} z_\ell z_{\ell+\ell'} \psi_{\bmk\nu}(\ell+\ell') ,
 \label{eq:Schrodinger}
\end{align}
with the kinetic energy $\epsilon_{\bmk} = -2t( \cos k_x a + \cos k_y a )$.
We define an order of the energy spectrum as $E_{\bmk,\nu=1} \leq E_{\bmk,\nu=2} \leq \cdots$.
The $\bmk$ dependence of $E_{\bmk\nu}$ is taken into account only through the function $\epsilon_\bmk$,
and thus the quasiparticle spectrum can be represented as $E_{\bmk\nu} = E_{\nu} (\epsilon_\bmk)$.
The factor $z_\ell^2$ represents a renormalization factor of the in-plane hopping arising from interactions.
The explicit form of $z_\ell$ is given by a function of $n_\ell$ and $d_\ell$ as \cite{Kotliar1986}
\begin{equation}
 z_\ell = 
  \frac{ \sqrt{(1-n_\ell+d_\ell^2)(n_\ell-2d_\ell^2)} + d_\ell \sqrt{n_\ell-2d_\ell^2} }
  {\sqrt{n_\ell(1-n_\ell/2)}} .
\end{equation}

\begin{table}
 \begin{tabular}{c|@{\hspace{15pt}}r@{\hspace{15pt}}r@{\hspace{15pt}}r}
  \toprule
  $n$ & 0 & 1 & 2 \\
  \colrule
  $\Gamma_\mathrm{c} (n+1/2)$ & -0.1389 & -0.0003 & $|...|<10^{-4}$ \\
  ${\tilde \Gamma} (n)$ & -1.6156 & -0.0071 &  $|...|<10^{-4}$ \\
  $\Delta(n)$ & -3.9003 & 0.0078 &  $|...|<10^{-4}$ \\
  \botrule 
 \end{tabular}
 \caption{Numerically estimated values of the correction terms in Eq.~(\ref{eq:bar_phi}).}
 \label{tab:correction}
\end{table}

The quantity $\bar{\phi}_\ell$ in Eq.~\eqref{eq:Schrodinger} represents the effective one-dimensional electrostatic potential
on the $\ell$-th layer including the additional term from the onsite repulsion $U_0$.
It can be obtained by integrating the $x$ and $y$ components in Eq.~(\ref{eq:phi}) as
\begin{align}
 \frac{\bar{\phi}_\ell}{\Ec}
  =&
  {\sum_{\ell'}}' 
 \Big\{
    2\pi \rhoC \left| \ell -  {\ell'}_{1/2} \right|
    + 2\pi \rhot \left| \ell -  \ell' \right|
 \Big\}
  \nonumber\\
 - & \sum_{\ell'} 2\pi n_{\ell'} \left| \ell -  \ell' \right| 
  \nonumber\\
 -& {\sum_{\ell'}}'
  \left\{
    \Gamma_\mathrm{c} ( \left| \ell - {\ell'}_{1/2} \right| ) \rhoC
   +\tilde{\Gamma} \left( \left| \ell - \ell' \right| \right) \rhot
  \right\}
   \nonumber\\
 +& \sum_{\ell'}
  \Delta \left( \left| \ell - \ell' \right| \right) n_{\ell'}
  +\frac{U_0}{4\Ec} n_\ell \delta_{\ell\ell'}
  ,
 \label{eq:bar_phi}
\end{align}
where ${\sum_{\ell'}}'=\sum_{-N/2+1 \le \ell' \le  N/2}$ and ${\ell'}_{1/2} = \ell'-1/2$.
The first and second terms represent the electrostatic potential of homogeneously charged infinite planes
located at $z/a=\ell'$ or $\ell'_{1/2}$.
The two subsequent terms originate from the difference between the electrostatic potential generated by the homogeneous planes and the one arising from the point-like charges $n_i$, $\rhoC$, and $\rhot$ located at discrete positions in the lattice model, and can be numerically calculated \cite{Wehrli2001} as presented in Table \ref{tab:correction}. The divergent contribution of each term in Eq.~(\ref{eq:phi}) is precisely canceled out due to the charge neutrality.

It is well known that the SBMF approach reproduces the Gutzwiller approximation 
in many situation on the mean-field level \cite{Kotliar1986}.
In the above approximation,
we treat the long-range Coulomb interaction $\phi_i$ 
and the part of the onsite interaction $U_0$ in the Hartree mean-field approximation,
while the remaining $\Ur=U-U_0$ is treated in the spirit of Gutzwiller approximation.
A self-consistent solution provides us with the properties of the coherent (low-energy) quasiparticle, which are renormalized by the electronic correlations.
For example, the in-plane quasiparticle velocity is given by
\begin{equation}
 {\bf v}_{\bmk\nu} = Z_{\bmk\nu} {\bm v}_{\bmk},
 \label{eq:vknu}
\end{equation}
with ${\bm v}_{\bmk} = {\bm \nabla} \epsilon_{\bmk} / \hbar$
and the renormalization amplitude
\begin{equation}
 Z_{\bmk\nu} = \frac{\partial E_{\bmk\nu}}{\partial \epsilon_{\bmk}}
  = \sum_\ell z_\ell^2 \psi_{\bmk\nu}(\ell)^2 \leq 1.
  \label{eq:Z}
\end{equation}
The coherent electron density is also directly accessible from the mean-field solution\cite{Ruegg2007}
\begin{equation}
n_{\rm coh}(\ell)=z_\ell^2n_\ell.
\label{eq:ncoh}
\end{equation}
Finally, we can also derive the retarded quasiparticle Green function as
\begin{equation}
 G_{\ell\ell'}^{(0)} (\bmk,\omega)
  =
  \sum_\nu \frac{\psi_{\bmk\nu}(\ell) \psi_{\bmk\nu}(\ell')}
  {\hbar \omega-E_{\bmk\nu}+i0^+}.
  \label{eq:G}
\end{equation}
All SBMF calculation shown in this paper are performed at $T=0$.
The free energy and its gradients are calculated in the thermodynamic limit $N_\parallel \rightarrow \infty$. Unless otherwise noted, we set the system size as $N=10$ and $M=30$, and impose open boundary conditions for the diagonalization of Eq.~(\ref{eq:Schrodinger}).
In the following calculation, we fix $\Ur=25t$.

%-----------------------------------------------------------------
\subsection{Transport coefficients}
%-----------------------------------------------------------------

\subsubsection{Transport distribution function}

Relying on the above $T=0$ SBMF approximation for the low-temperature electric properties \cite{Ruegg2007,Ruegg2008},
we derive the transport coefficients within the linear-response Kubo formalism \cite{Ruegg2008}.
In this work we focus on the in-plane longitudinal dc electrical conductivity $\sigma$ and the Seebeck coefficient $S$, which are expressed as
\begin{align} 
 \sigma =& L^{(0)},  &
 S =& - \frac{1}{eT} \frac{L^{(1)}}{L^{(0)}}.   \label{eq:total}
\end{align}
Here, the Seebeck coefficient was obtained by applying the Jonson-Mahan theorem \cite{Jonson1980}.
The functions $L^{(n)}$ are defined as
\begin{equation}
 L^{(n)} = \int dE \left(-\frac{\partial f(E)}{\partial E} \right)
  E^n \Phi(E),
  \label{eq:L}
\end{equation}
with $f(E)=(1+e^{E/\kB T})^{-1}$ and the transport distribution function,
\begin{equation}
 \Phi(E) = \frac{2\pi e^2 \hbar }{ N_\parallel} \sum_\bmk
  \left( v_\bmk^x \right)^2 \ \mathrm{Tr} \left[ \hat{A} (\bmk,E)^2 \right].
  \label{eq:TDF}
\end{equation}
Here, $ \hat{A} (\bmk,E) $ is the spectral density matrix of quasiparticles,
$ \hat{A} (\bmk,E) = -\frac{1}{\pi} \mathrm{Im} \ \hat{G} (\bmk,E) $,
with the retarded Green function $\hat{G} (\bmk,E)$, which is derived in the following section.

\subsubsection{Impurity scattering}

We derive the spectral density matrix $ \hat{A} (\bmk,E) $ under the assumption
that the dominant relaxation mechanism at low temperatures is elastic scattering by impurities or vacancies.
For this purpose, we add a short-range impurity Hamiltonian to our model,
\begin{equation}
 H_\mathrm{imp} = V_0 \sum_{\sigma,i'} c_{i'\sigma}^\dagger c_{i'\sigma},
\end{equation}
where the label $i'$ denotes impurity sites.
We assume a dilute impurity concentration $c_\mathrm{imp}$ which is on average independent of the layer index $\ell$.
In this case, the self-energy of the quasiparticles due to impurity scattering, $\Sigma_{\nu\nu'}^\mathrm{imp} (\bmk,\omega)$,  can be obtained from the T-matrix approximation using the quasiparticle Green function of the pure system (\ref{eq:G}).
If we neglect the real part of $G_{\ell\ell'}^{(0)} (\bmk,\omega)$ and the non-diagonal term of the self energy, 
the imaginary part of the total self energy is given as
\begin{equation}
 \gamma_{\bmk\nu} (\omega) 
  \equiv
  - \mathrm{Im} \left[ \Sigma_{\nu\nu} (\bmk,\omega) \right]
  = \gamma_{\bmk\nu}^\mathrm{imp} (\omega) + \gamma'  .
  \label{eq:gamma}
\end{equation}
$\gamma_{\bmk\nu}^\mathrm{imp} (\omega)=-{\rm Im}\left[\Sigma_{\nu\nu}^\mathrm{imp} (\bmk,\omega)\right]$ originates from the impurity scattering  
and is given by
\begin{equation}
 \gamma_{\bmk\nu}^\mathrm{imp} (\omega)
 =
  c_\mathrm{imp} V_0  \sum_\ell 
  \frac{\psi_{\bmk\nu}(\ell)^2  \left\{ \pi V_0 \rho_\ell(\omega) \right\} }
  {1 + \left\{ \pi V_0 \rho_\ell(\omega) \right\}^2} ,
 \label{eq:gamma_imp}
\end{equation}
with the layer-resolved quasiparticle spectral density of the pure system,
\begin{equation}
 \rho_\ell(\omega) 
  = \frac{1}{N_\parallel} \sum_\bmk 
  \left( - \frac{1}{\pi} \mathrm{Im} G_{\ell\ell}^{(0)} (\bmk, \omega)  \right) .
%  = \frac{1}{N_\parallel} \sum_{\bmk \nu} \psi_{\bmk\nu}(\ell)^2 
%  \delta ( \hbar \omega-E_{\bmk\nu}).
\end{equation}
The additional quantity $\gamma'$ in Eq.~(\ref{eq:gamma}) is introduced to include contributions from other factors,
e.g., electron-phonon couplings, high-energy corrections due to electron-electron interactions, or lattice disorder,
and is assumed to be constant for simplicity.
Hereafter, we ignore the real part of the self energy:
note that the real part of $\Sigma_{\nu\nu'}^\mathrm{imp} (\bmk,\omega)$
has no contributions to $\mathrm{Tr} [ \hat{A} (\bmk,E)^2 ]$  in the limit of $c_\mathrm{imp} \to 0$.
In this case, we obtain
\begin{equation}
 \mathrm{Tr} \left[ \hat{A} (\bmk,\omega)^2 \right] 
  =
  \frac{1}{\pi \hbar} \sum_\nu Z_{\bmk\nu}^2 \tau_{\bmk\nu}(\omega) 
  \delta( \hbar \omega - E_{\bmk\nu}),
\end{equation}
for the dilute impurities in the limit of $c_\mathrm{imp} \to 0$
with the relaxation time $\tau_{\bmk\nu}(\omega) = \hbar /2\gamma_{\bmk\nu}(\omega)$.

\subsubsection{Total and resolved transport coefficients}

By introducing the weighted density of states
\begin{equation}
 \mathcal{N} (\epsilon) = \frac{1}{(2\pi)^2} \int d\bmk
  \left| {\bm v}_\bmk  \right|^2 
  \delta(\epsilon-\epsilon_\bmk),
\end{equation}
we can finally obtain the transport distribution function (\ref{eq:TDF}) 
in the following form:
\begin{equation}
 \Phi(\omega) = \frac{e^2}{\hbar} \sum_\nu 
  Z_\nu^* (\omega) \tau_\nu^* (\omega) \mathcal{N}(\epsilon_{\nu}^*(\omega)) .
  \label{eq:TDF_total}
\end{equation}
Here $Z_\nu^* (\omega)$, $\tau_\nu^* (\omega)$ and $\epsilon_{\nu}^* (\omega)$ are
values of $Z_{\bmk\nu}$, $\tau_{\bmk\nu}(\omega)$ and $\epsilon_{\bmk}$
for $\bmk = \bmk_{\nu}^* (\omega)$, respectively,
where $ \bmk_{\nu}^* (\omega) $ is determined by 
$ E_{\bmk_{\nu}^*(\omega), \nu} = E_{\nu}(\epsilon_{\nu}^* (\omega)) = \hbar \omega $.
In the derivation of Eq.~(\ref{eq:TDF_total}), we assume nondegenerate subbands.

The resultant total transport distribution function (\ref{eq:TDF_total}) is a sum of contributions from each subband, 
$ \Phi(\omega) = \sum_\nu \Phi_\nu (\omega) $,
where 
\begin{equation}
 \Phi_\nu (\omega) =  \frac{e^2}{\hbar} 
  Z_\nu^* (\omega) \tau_\nu^* (\omega) \mathcal{N}(\epsilon_{\nu}^*(\omega)) .
  \label{eq:TDF_band}
\end{equation}
With this function, we can define the band-resolved conductivity and Seebeck coefficients as
\cite{Ruegg2008}
\begin{align}
 \sigma_\nu =& L_\nu^{(0)},     &
 S_\nu =& - \frac{1}{eT} \frac{L_\nu^{(1)}}{L_\nu^{(0)}}, 
\end{align}
where $L_\nu^{(n)}$ is the Fermi integrals over the band-resolved transport distribution function $\Phi_\nu (\omega)$ in the same way as Eq.~(\ref{eq:L}).
% %
% \begin{equation}
%  L_\nu^{(n)} = \int dE \left(-\frac{\partial f(E)}{\partial E} \right)
%   E^n \Phi_\nu(E).
%   \label{eq:L_band}
% \end{equation}
% %
These values represent the contributions from each subband to the total transports,
$\sigma = \sum_\nu \sigma_\nu$ and $S = \sum_\nu S_\nu \sigma_\nu / \sigma$.

On the other hand,
we can also define the layer-resolved transport distribution function $\Phi_\ell(\omega)$
from Eq.~(\ref{eq:TDF}) as
\begin{align}
 \Phi_\ell(E) \equiv& 
 \frac{2\pi e^2 \hbar}{N_\parallel} \sum_\bmk
  \left( v_\bmk^x \right)^2 \left[ \hat{A} (\bmk,E)^2 \right]_{\ell\ell} 
 \nonumber\\
 =&
\frac{e^2}{\hbar}  z_\ell^2 \sum_\nu
  \psi_{{\bmk_\nu^*(\omega)},\nu} (\ell)^2 \tau_\nu^* (\omega) \mathcal{N}(\epsilon_{\nu}^*(\omega)) ,
  \label{eq:TDF_layer}
\end{align}
which is obtained from the same calculations as Eq.~(\ref{eq:TDF_total}),
and satisfies $ \Phi(\omega) = \sum_\ell \Phi_\ell (\omega) $.
With this function, we can define the layer-resolved conductivity and Seebeck coefficients as
\begin{align}
 \sigma_\ell =& L_\ell^{(0)},  &
 S_\ell =& - \frac{1}{eT} \frac{L_\ell^{(1)}}{L_\ell^{(0)}}, 
\end{align}
where $L_\ell^{(n)}$ is obtained from $\Phi_\ell(\omega)$.
These values represent the contributions from each layer to the total transport,
$\sigma = \sum_\ell \sigma_\ell$ and $S = \sum_\ell S_\ell \sigma_\ell / \sigma$.

All the results presented in this paper are obtained using the following parameters: $\kB T=0.01t$,
the impurity concentration $c_\mathrm{imp}=0.1$,
the impurity scattering $V_0=0.3 t$,
and $\gamma'=0.001$, 
which gives $\gamma^\mathrm{imp}_{\bmk\nu}(\omega) > \gamma'$ except in the vicinity of the phase transition,
as shown in Sec.~\ref{sec:MI_transition}.

%=================================================================
\section{Ground-state properties}
\label{sec:ground_state}
%=================================================================

%-----------------------------------------------------------------
\subsection{Charge distribution and electrostatic potential}
%-----------------------------------------------------------------

\begin{figure}
 \centering
 \includegraphics[]{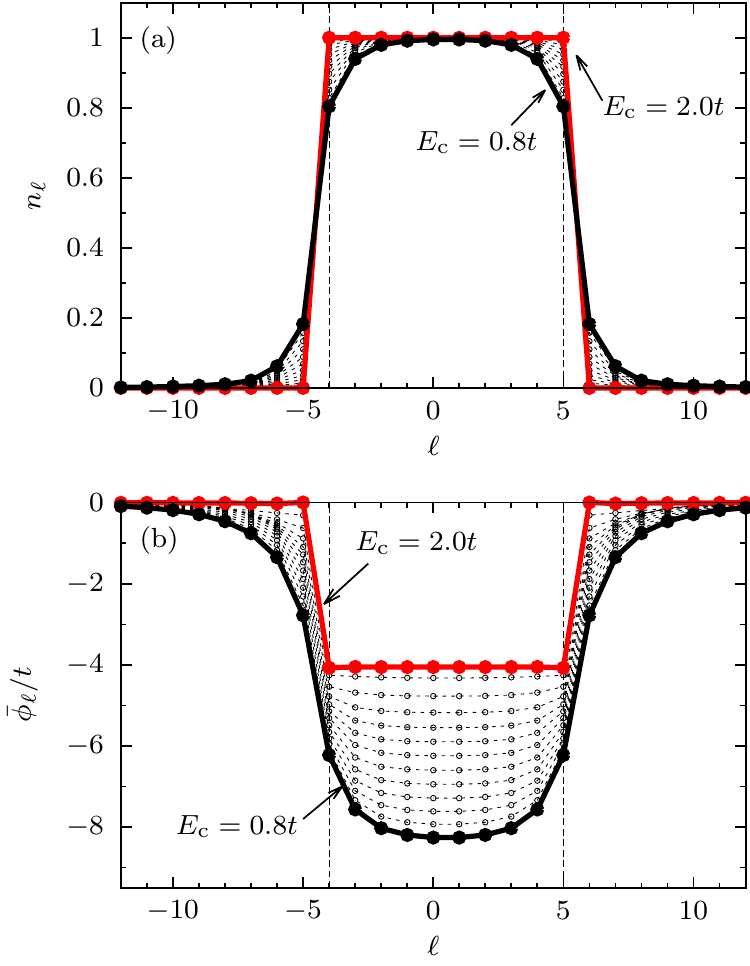}
 \caption{
 (Color online)
 (a) The electronic charge distribution $n_\ell$ and
 (b) the effective 1D electrostatic potential ${\bar \phi}_\ell$
 for $\Ur=25t$ with $\Ec=0.8t,\cdots,2.0t$ around the MI region in the $\rhot=1$ structure.
 The dashed lines indicate the position of the interfaces, $\ell=-4$ and $5$.
 }
 \label{fig:n-phi}
\end{figure}

For the emergence of an interfacial metallic state between a Mott and a band insulator,
the electronic charge distribution around the interface is of paramount importance.
Deviations of the density from that of the homogeneous systems, i.e., $n=0$ in the BI and $n=1$ in the MI material,
directly leads to electronic conductivity.
In Fig.~\ref{fig:n-phi}(a), we show the layer-resolved electronic charge density $n_\ell$
for various values of $\Ec$ in the $\rhot=1$ nonpolar heterostructure. We fix $\Ur=25t$ for the following discussion.
The electrons reside in the central region where the positive charges are localized,
and realize the bulk densities $n_\ell \sim 1$ or $n_\ell \sim 0$ away from the interface.
However, for the plausible value $\Ec=0.8t$,
the distribution is broad around the interface with $n_{\ell=5} \sim 0.8$ and  $n_{\ell=6} \sim 0.2$.
This smooth distribution changes with increasing $\Ec$, and becomes completely ``sharp'' for $\Ec \geq 1.96t$: 
all the layers are either at a density $n_\ell=0$ or $1$, even at the interface.

The screened electrostatic potential follows the behavior of the electronic charge distribution.
In Fig.~\ref{fig:n-phi}(b) we show ${\bar \phi}_\ell$, which represents the electrostatic potential on the $\ell$-th layer including the contribution from $U_0$, see Eq.~\eqref{eq:bar_phi}.
For small $\Ec$, the smooth charge distribution results in interfacial dipole moments formed by positively- and negatively-charged planes around the interface in the MI and BI regions, respectively.
These interfacial dipoles lower the electrostatic potential in the center region due to the first two terms of Eq.~(\ref{eq:bar_phi}).
When $\Ec>1.96t$ all the layers are neutral, and only the last three terms in Eq.~\eqref{eq:bar_phi} contribute. This leads to a piecewise constant electrostatic potential as seen in Fig.~\ref{fig:n-phi}(b).

The dependence of $n_\ell$ and ${\bar \phi}_\ell$ on $\Ec$ (or the dielectric constant $\varepsilon$) is in agreement with the expectation from the Thomas-Fermi screening length Eq.~\eqref{eq:TF}: like the spread of the itinerant electron density $n_\ell$ around the localized positive background charges, $\lambda_{\rm TF}$ is a decreasing function of $\Ec$. For large $\Ec$ (small $\varepsilon$), charge fluctuations are strongly suppressed, and both $\lambda_{\rm TF}$ as well as the effective Bohr radius Eq.~\eqref{eq:a_B} approach the size of the lattice constant. Only in this limit we recover the ionic picture and an insulating interface.\cite{Ohtomo2004,Nakagawa2006,Bristowe:2014}

%-----------------------------------------------------------------
\subsection{Localization and renormalization}
\label{sec:loc_and_ren}
%-----------------------------------------------------------------
%
\begin{figure}
 \centering
 \includegraphics[]{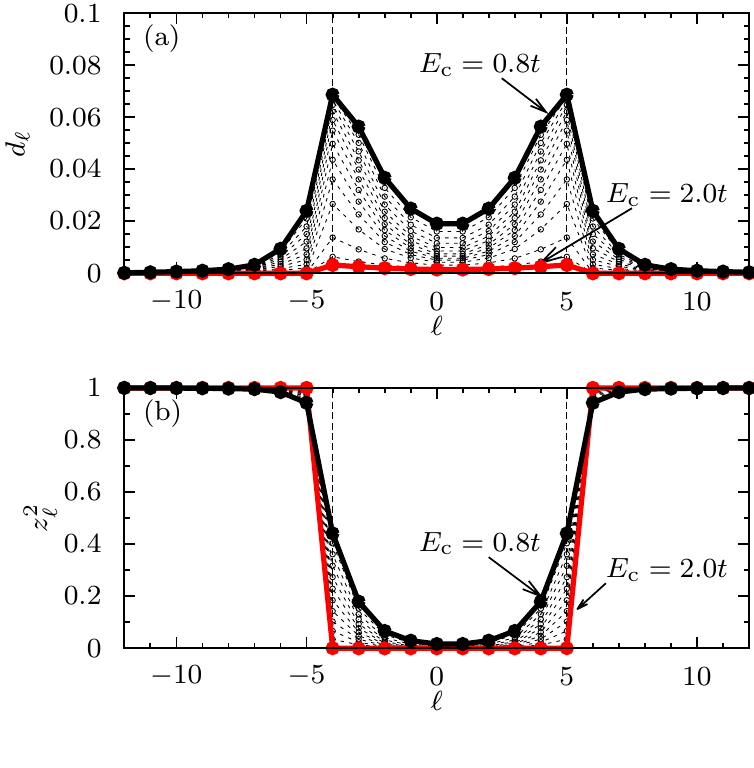}
 \caption{
 (Color online)
 (a) The amplitude of the double occupancy $d_\ell$ and
 (b) the in-plane-hopping renormalization factor $z_\ell^2$
 for $\Ur=25t$ with $\Ec=0.8t,\cdots,2.0t$ around the MI region in the $\rhot=1$ structure.
 The dashed lines indicate the position of the interfaces, $\ell=-4$ and $5$.
 }
 \label{fig:d-z}
\end{figure}
The electrons in the central region are localized due to the strong onsite Coulomb repulsion $\Ur$.
Within the SBMF method, the localization is apparent from the amplitude of the double occupancy $d_\ell$ and the in-plane-hopping renormalization factor $z_\ell^2$, which are presented in Fig.~\ref{fig:d-z} for various $\Ec$ in the $\rhot=1$ heterostructure.
It is obvious that they are reduced by increasing $\Ur$ as already studied in the $\rhot=0$ polar/nonpolar heterostructure\cite{Ruegg2007}.
Again, the dependence on $\Ec$ can be understood from the behavior of the electron density $n_\ell$: localization effects are larger if $n_\ell\rightarrow 1$.
The value of the double occupancy at the interface for $\Ec=0.8t$ is  $d_{\ell=5}=0.068$, which is similar to the value found in the $\rhot=0$ polar/nonpolar heterostructure \cite{Ruegg2007}.
With increasing $\Ec$, the double occupancy tends to zero in the whole system, which is accompanied by the change of the electron densities to the step-like distribution $n_\ell \rightarrow 0$ or $1$. As a consequence, the in-plane-hopping renormalization factor $z_\ell^2$ also goes to zero in the central region,
as shown in Fig.~\ref{fig:d-z}(b). Note that the amount $z_\ell^2$ also gives the mass renormalization in the $\ell$-th layer, $z_\ell^2 = m/m_\ell^*$.
These results demonstrate Mott's metal-insulator transition in the center material tuned by the dielectric constant (or $\Ec$).

\begin{figure}
 \centering
 \includegraphics[]{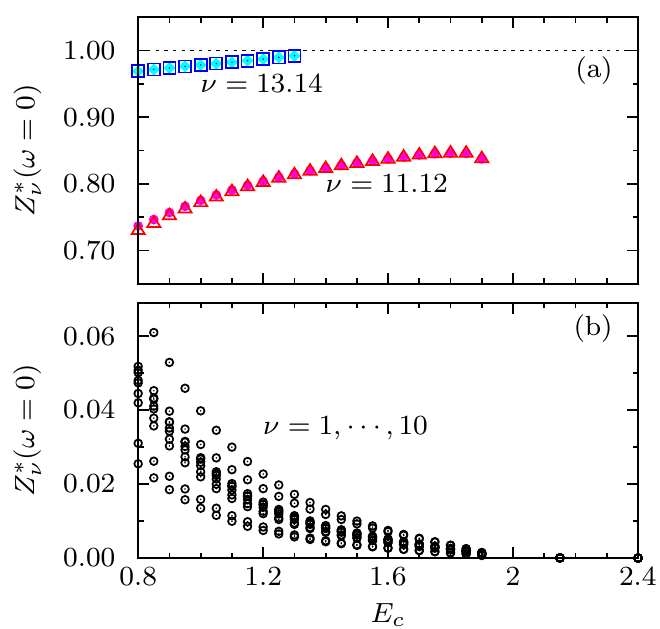}
 \caption{
 (Color online)
 The renormalization amplitude of the quasiparticle velocity $Z_{\bmk\nu}$
 for $\bmk$ satisfying $E_{\bmk\nu}=\hbar\omega=0$, i.e., $Z_\nu^*(\omega=0)$,
 as a function of $\Ec$ for $\Ur=25t$ in the $\rhot=1$ structure.
 }
 \label{fig:Z}
\end{figure}

\begin{figure}
 \centering
 \includegraphics[]{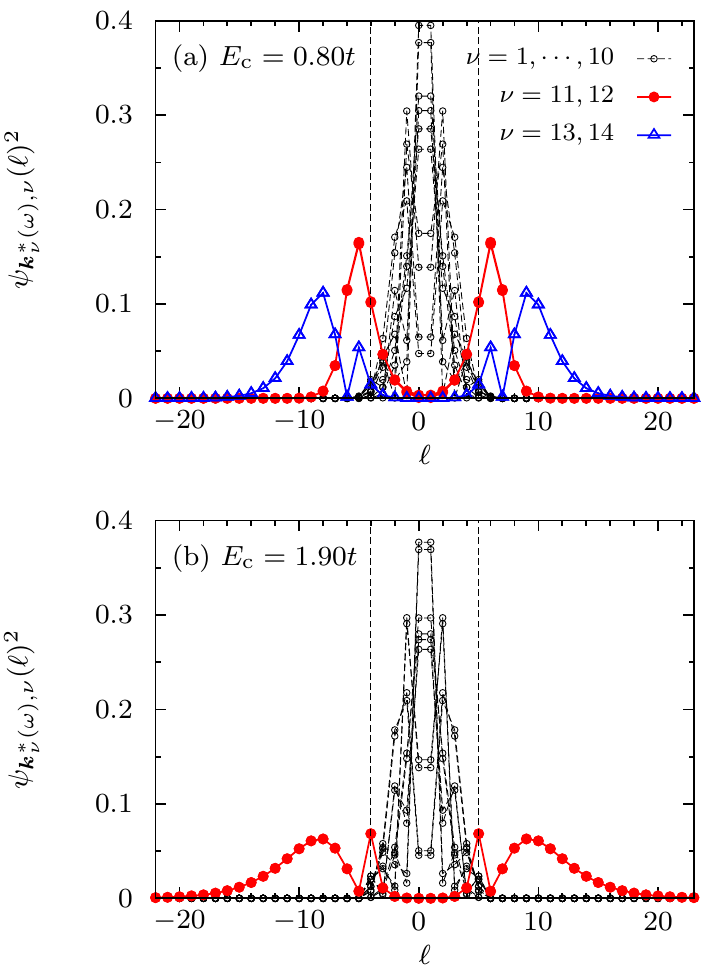}
 \caption{
 (Color online) Square of the wave function at the Fermi energy $\psi_{\bmk_\nu^*(\omega=0),\nu}(\ell)^2$ for
 (a) $\Ec=0.80t$ and (b) $\Ec=1.90t$, and $\Ur=25t$ and $\rhot=1$.
 The dashed lines indicate the position of the interfaces, $\ell=-4$ and $5$.
 }
 \label{fig:psi2}
\end{figure}

The renormalization effects are also apparent in the quasiparticle velocity ${\bf v}_{\bmk\nu}=Z_{\bmk\nu}{\bm v}_{\bmk}$, see Eq.~\eqref{eq:vknu}.
We show its renormalization factor $Z_{\bmk\nu}$ as a function of $\Ec$ for the $\rhot=1$ heterostructure in Fig.~\ref{fig:Z}.
We present the value at the Fermi energy, $Z_\nu^*(\omega=0)$,
which is one of the key quantities to determine the transport coefficients [see Eqs.~\eqref{eq:TDF_band} and \eqref{eq:TDF_layer}].
High-energy bands with the index $\nu>15$ lie above the Fermi energy for $\Ec > 0.8t$, and $Z_\nu^*(\omega=0)$ is undefined (however, $Z_{\bmk\nu}=1$ for $\nu>15$). The wave functions $\psi_{\bmk\nu}(\ell)$ of quasiparticles belonging to the low-energy bands $\nu \leq 10$ have spatial weights only in the central layers as shown in Fig.~\ref{fig:psi2}. Consequently the velocities derived from these bands are strongly renormalized and tend to zero with increasing $\Ec$.
On the other hand, the quasiparticles of the intermediate bands $\nu=11,\cdots,14$ are located in the vicinity of the interface.
Since the wave functions of the higher-energy bands $\nu=13$ and $14$ extend toward the BI area [Fig.~\ref{fig:psi2}(a)],
the value of $Z_{\nu}^*(0)$ is close to 1 indicating an exceedingly weak renormalization.
Upon reaching $Z_\nu^*(0)=1$ at $\Ec \sim 1.4t$,
these two bands become depopulated ($E_{\bmk\nu}>0$). The lower bands $\nu=11$ and $12$ become depopulated only at $\Ec \sim 1.96t$, which is the metal-insulator transition point.
Although these bands keep large velocities compared with the bands $\nu=1,\cdots,10$,
the renormalization factors do not reach 1, because their wave functions penetrate the MI region two or three layers deep (Fig.~\ref{fig:psi2}).
These results illustrate the correspondence between bands and spatial regions of the heterostructure \cite{Ruegg2009} as introduced in Fig.~\ref{fig:sigma_rhot0-1}(b): The $\nu \leq 10$ bands correspond to the MI region, and $\nu=11,12$ to the interface.
The others can be assigned to the BI region with the dominant bands $\nu=13$ and $14$.

%-----------------------------------------------------------------
\subsection{The metal-insulator transition at the interface}
%-----------------------------------------------------------------

\begin{figure}
 \centering
 \includegraphics[]{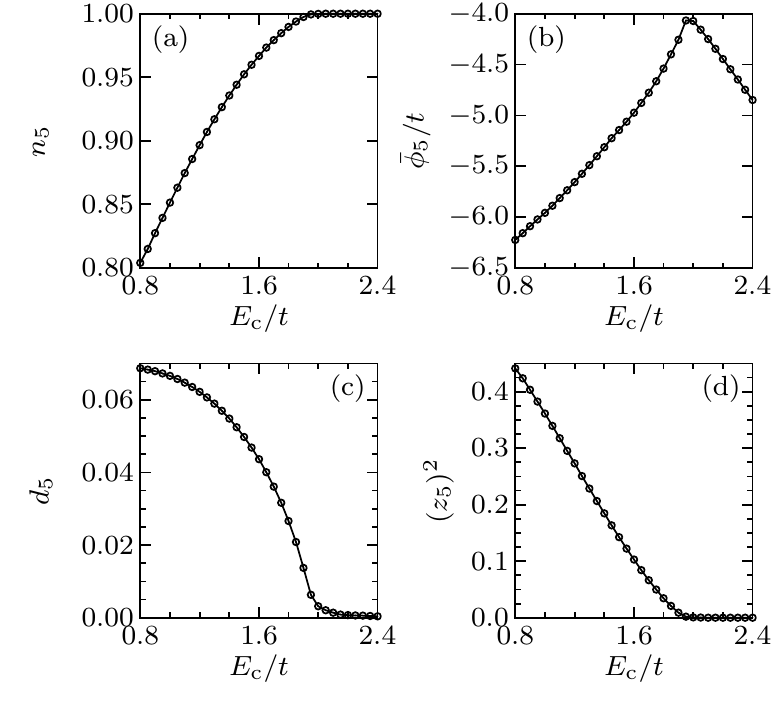}
 \caption{
 The $\ell=5$ interfacial quantities as a function of $\Ec$ for $\Ur=25t$ in the $\rhot=1$ structure.
 (a) The electronic charge density $n_\ell$,
 (b) the effective 1D electrostatic potential ${\bar \phi}_\ell$,
 (c) the amplitude of the double occupancy $d_\ell$, and
 (d) the in-plane-hopping renormalization factor $z_\ell^2$.
 }
 \label{fig:npdz}
\end{figure}

We summarize the ground-state properties in Fig.~\ref{fig:npdz} as a function of $\Ec$ with focus on the interface layer, $\ell=5$.
At the transition point $\Ec=1.96t$, the charge density at the interface reaches $n_{\ell=5}=1$ accompanied by $d_{\ell=5} = 0$ and $z_{\ell=5}=0$. It indicates the full occupation of the lower Hubbard band and the Mott transition at the interface.
The transition is captured most clearly by the electrostatic potential ${\bar \phi}_{\ell=5}$ presented in Fig.~\ref{fig:npdz}(b).
Below the transition, $\Ec<1.96t$,
${\bar \phi}_{\ell=5}$ is dominated by contributions from infinite planes with a non-vanishing net charge,
resulting in a increasing function of $\Ec$.
After the transition, all planes are neutral, and the above contributions drop down.
The remaining correction terms in Eq.~(\ref{eq:bar_phi}) make ${\bar \phi}_\ell$ the linear decreasing function of $\Ec$.
Consequently, the electrostatic potential ${\bar \phi}_\ell$ on the interface has a kink at the transition point.

%=================================================================
\section{Transport properties}
\label{sec:MI_transition}
%=================================================================

%-----------------------------------------------------------------
\subsection{Relaxation time}
%-----------------------------------------------------------------

The transport distribution functions (\ref{eq:TDF_total}), (\ref{eq:TDF_band}) and (\ref{eq:TDF_layer}) consist of three factors: 
The renormalization amplitude of the quasiparticle velocity $Z_\nu^*(\omega)$ or its components $z_\ell^2 \psi_{\bmk_\nu^*(\omega),\nu}(\ell)^2$, the relaxation time $\tau_\nu^*(\omega)$, and the weighted density of state ${\mathcal N}(\epsilon^*_\nu(\omega))$.
The renormalization amplitude satisfies $Z_\nu^*(\omega) \leq 1$, see Fig.~\ref{fig:Z}.
The weighted density of state ${\mathcal N}(\epsilon)$, which corresponds to a square mean value of velocity in the noninteracting 2D homogeneous system, can be obtained precisely in terms of the complete elliptic integrals of the first and second kinds
and ${\mathcal N}(\epsilon) \leq 8t/(\pi\hbar)^2$.

\begin{figure}
 \centering
 \includegraphics[]{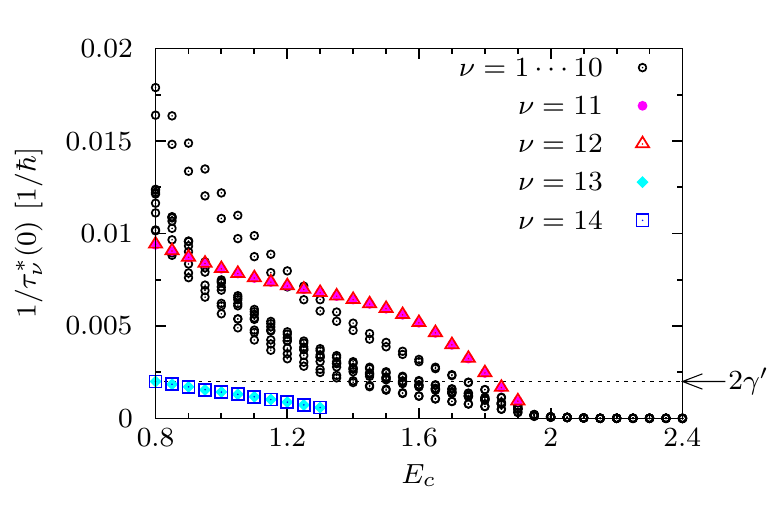}
 \caption{
 (Color online) 
 The inverse of the relaxation time at the Fermi energy, $\tau_\nu^*(\omega=0)^{-1}$,
 for $\Ur=25t$, $\rhot=1$ and $\gamma'=0$,
 which means $1/\tau_\nu^*(0) = 2\gamma_{\bmk\nu}^\mathrm{imp}(0)/\hbar$ in this figure.
 }
 \label{fig:tau-1}
\end{figure}

Figure \ref{fig:tau-1} shows the inverse of the relaxation time at the Fermi energy, $1/\tau_\nu^*(0)$,
in the $\rhot=1$ nonpolar heterostructure.
Only in this figure, we set $\gamma'=0$ to focus on impurity effects.
Different behaviors of the relaxation time between the lower $\nu \leq 10$ and higher $\nu=11,\cdots,14$ energy bands are determined by a parameter $\pi V_0 \rho_\ell(0)$.
For $\nu \leq 10$, the wave functions $\psi_{\bmk\nu}(\ell)$ on the Fermi energy are located in the MI region as shown in Fig.~\ref{fig:psi2}, where the quasiparticle spectral densities $\rho_\ell(\omega)$ are strongly confined around $\omega=0$\cite{Ruegg2007}.
Since this confinement becomes stronger in association with the renormalization effect,
$\rho_\ell(0)$ are increasing functions of $\Ec$ and $\pi V_0 \rho_\ell(0) \gg 1$ for $\ell$ belonging to the MI region.
Thus the contributions to $\gamma_{\bmk_\nu^*(0),\nu}^\mathrm{imp}(0)$ for $\nu \leq 10$ from each layer gradually decrease with increasing $\Ec$.
On the other hand, the higher-energy bands $\nu=11,\cdots,14$ correspond to the outside region of the heterostructure, and $\pi V_0 \rho_\ell(0)$ take on small values even for $\Ec=0.80t$.
Since the values of $\rho_\ell(0)$ in the BI region are reduced and tend to zero with losing the electrons,
$\gamma_{\bmk_\nu^*(0),\nu}^\mathrm{imp}(0)$ decrease linearly with increasing $\Ec$ for $\nu=11,\cdots,14$.

%-----------------------------------------------------------------
\subsection{Conductivity}
%-----------------------------------------------------------------

\begin{figure}
 \centering
 \includegraphics[]{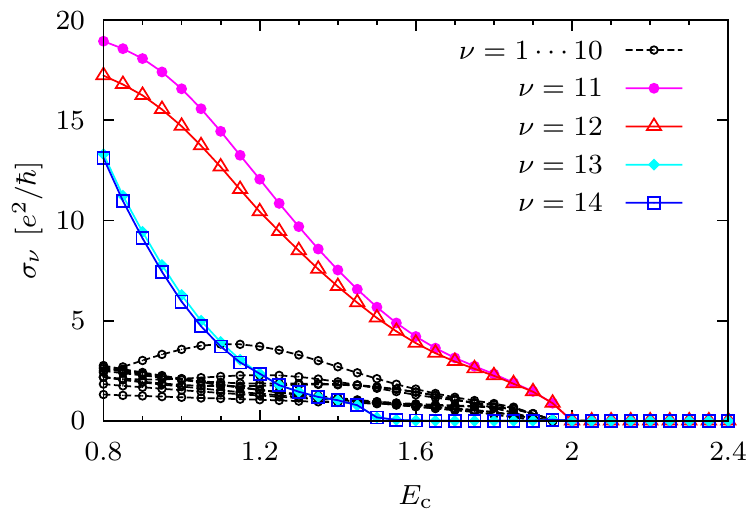}
 \caption{
 (Color online) 
 The band-resolved conductivity $\sigma_\nu$ for $\Ur=25t$ and $\rhot=1$.
 }
 \label{fig:cond_band}
\end{figure}

In Fig.~\ref{fig:cond_band} we show the band-resolved conductivities as a function of $\Ec$.
The conductivities for the lower bands $\nu \leq 10$ are suppressed due to the strong renormalization of the velocity
$ {\bf v}_{\bmk\nu} = Z_{\bmk\nu} {\bm v}_{\bmk}$.
The higher-energy bands $\nu=13$ and $14$ have a large relaxation time $\tau_\nu^*(0)$ and velocities ${\bf v}_{\bmk\nu}$, but the weighted density of states ${\cal N}(\epsilon_\nu^*(0))$ are small (since $\epsilon_\nu^*(0)$ are close to $\epsilon \sim -4.0t$) resulting in a small contribution to the total conductivity with increasing $\Ec$.
Consequently, dominant contributions to the conductivity are given by quasiparticles on the bands $\nu=11$ and $12$.
These observations are consistent with the layer-resolved conductivities presented in Fig.~\ref{fig:LayerTransports}(a):
The distributions of the wave functions for the dominant $\nu=11,12$ and second-dominant $\nu=13,14$ energy bands presented in Fig.~\ref{fig:psi2} reproduce that of the layer-resolved conductivity shown in Fig.~\ref{fig:LayerTransports}(a). It is also confirmed in Fig.~\ref{fig:sigma_rhot0-1} that the total conductivity of the BI region slightly changes around $\Ec=1.4t$ due to the drop of the contributions from the depleted bands $\nu=13$ and $14$.
We note that if only the impurity scattering is taken into account to the self energy (\ref{eq:gamma}),
the relaxation time $\tau_\nu^*(0)$ diverges upon reaching $\Ec \rightarrow 1.96t$ (Fig.~\ref{fig:tau-1}), which leads to a unphysical rise in the conductivity.
In other words, the phenomenological constant $\gamma'$ is crucial for the transport in the vicinity of the transition because it becomes the dominant scale.

From the relation between the parameter $\pi V_0 \rho_\ell(0)$ and $1/\tau_\nu^*(0)$, cf.~Eq.~\eqref{eq:gamma_imp}
, we can also infer the dependence of the transport coefficients on the impurity potential $V_0$.
For the quasiparticles on the $\nu \leq 10$ lower energy bands,
the corresponding $\rho_\ell(0)$ are quite large so that the relaxation times $\tau_\nu^*(0)$ 
are expected to be unaffected largely by $V_0$.
In contrast, $\tau_\nu^*(0)$ for $\nu=11,\cdots,14$ consists of the contributions from the outside layers with small $\rho_\ell(0)$,
where the parameter satisfies $\pi V_0 \rho_\ell (0) <1$.
Thus $\tau_\nu^*(0)$ can be varied significantly by $V_0$ for these bands.
Accordingly, the conductivity is expected to be highly dependent on the impurity scattering $V_0$ in the BI region,
but not in the center MI region.
In this manner, the parameters introduced to calculate the transport coefficients could change the layer dependence 
and the values of the coefficients quantitatively, but they do not intrinsically participate in the determination of the transition point.

%-----------------------------------------------------------------
\subsection{Seebeck coefficient}
%-----------------------------------------------------------------

\begin{figure}
 \centering
 \includegraphics[]{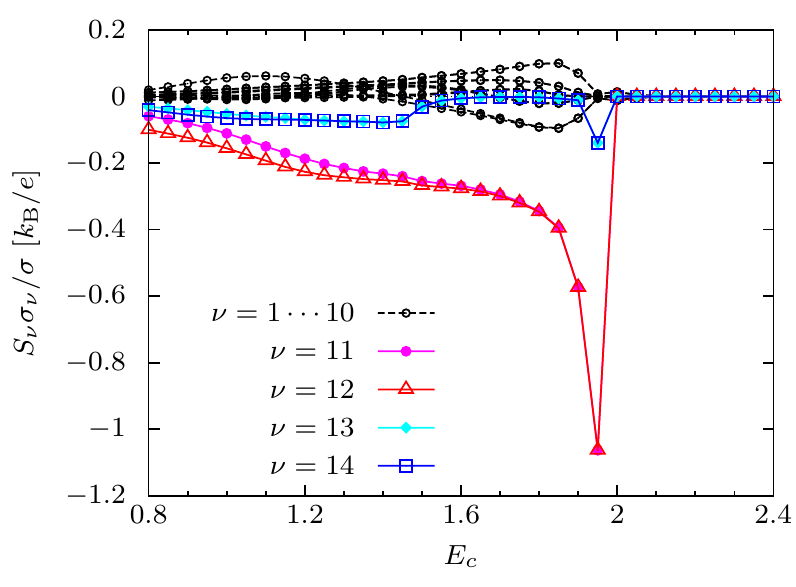}
 \caption{
 (Color online)
 The band-resolved contribution to the total Seebeck coefficient, $S_\nu \sigma_\nu/\sigma$
 for $\Ur=25t$ and $\rhot=1$.
 }
 \label{fig:Seebeck_band}
\end{figure}

The quasiparticles on the energy bands $\nu=11$ and $12$ also dominate the Seebeck coefficient.
We show the dependence of the band-resolved Seebeck coefficients on $\Ec$ in Fig.~\ref{fig:Seebeck_band}.
The contributions from these two bands are strongly enhanced at the metal-insulator transition point $\Ec=1.96t$,
where the chemical potential approaches the lower edge of these bands. Note that although the band edges of $\nu=13$ and $14$ similarly move towards the chemical potential at $\Ec=1.4t$,
the contributions of these bands to the Seebeck coefficient are inhibited by $\gamma'$.
For $\Ec>1.4t$, the $\nu=13$ and $14$ bands lose the weights on the Fermi energy, and thus their Seebeck coefficients become significantly small. 
However, finite values of the Seebeck coefficients for $\nu=13$ and $14$ are observed again in the vicinity of the phase transition,
because the band edges moves down to the chemical potential.
The quasiparticles on the $\nu=11$ and $12$ bands are located in the vicinity of the interface (Fig.~\ref{fig:psi2}), and thus the enhancement of the Seebeck coefficient is spatially confined to this region, as presented in Fig.~\ref{fig:LayerTransports}.

%=================================================================
\section{Conclusion}
\label{sec:conclusion}
%=================================================================

In this work, we have examined interfacial metallicity in a nonpolar/nonpolar heterostructure (i.e.~without a polar discontinuity) when a Mott and a band insulator are combined in a sandwich structure. Using the SBMF theory in combination with the relaxation-time approximation, we studied the influence of the dielectric constant on the electronic charge distribution and the in-plane transport. Most importantly, we find a metal-insulator transition tunable by the dielectric permittivity. 

If the dielectric constant is above a critical value (which depends on the geometry and the onsite Hubbard interaction), a high conductivity was obtained on several layers in the vicinity of the interface. Interestingly, the characteristics of this phase bear a great similarity to the one found in polar/nonpolar structures with a similar dielectric constant. If the screening length decreases, the conductivity gradually tends to zero, while the Seebeck coefficient shows a strong enhancement in the vicinity of the interface. Finally, when the dielectric constant falls below the critical value, the system undergoes the metal-insulator transition. The existence of such a metal-insulator transition in the nonpolar/nonpolar structure is in sharp contrast to the case of the polar/nonpolar heterostructure, in which at least the interface remains metallic for all values of the dielectric constant.

Nevertheless, our results imply that the polar discontinuity is not an indispensable ingredient to obtain interfacial metallicity in MI/BI heterostructures. Furthermore, the parameter range for metallicity (as obtained from the SBMF approximation) definitely lies in a physical regime. In fact, the condition $\Ec \equiv e^2/(\varepsilon a) < 1.9t$ for metallicity is well above earlier estimates of $\Ec=0.8t$ for the LTO/STO heterostructure.\cite{Ohtomo2002,Okamoto2004_PRB70} We also note that the dielectric constant of certain insulating perovskites (such as STO) strongly depend on temperature and electric field. The metal-insulator transition discussed in this paper therefore might be externally tunable.

Intriguingly, it is theoretically reported that a $\mathrm{SrNbO}_3$/STO heterostructure is expected to show metallic transport properties without polar discontinuity\cite{Zhong2012}. In addition, we expect that our theoretical setting for a heterostructure without a polar discontinuity has some relevance for perovskite-type MI/BI heterostructures grown along the [110] direction, which also show no polar discontinuity across the interface. In this context, it is interesting to note a recent 
experimental finding of emergent metallicity on the (110) LAO/STO interface\cite{Annadi:2013}. In contrast to the model discussed in this paper, this system involves two different BIs. However, we expect that at least the metallic phase can in principle be compared to the one we found here, although the origin of the metallicity might be different in the actual experiment and beyond the scope of our model.\cite{Annadi:2013} We are not aware of an experimental study of a nonpolar BI/MI heterostructure but the (110) LTO/STO structure would offer an experimentally accessible system.

%
%
%

%=================================================================
%=================================================================

\begin{acknowledgments}
We would like to thank G.~Sawatsky, S.~Pilgram and D.~M\"uller for stimulating discussions. Y.O. is grateful for the hospitality of the Pauli Center for Theoretical Studies of ETH Zurich. We are also grateful for financial support of the Swiss National Science Foundation through the Sinergia TEO and the Ambizione program (A.R.).
\end{acknowledgments}

%=================================================================
%=================================================================

% \cite{Staudt:2000}

\bibliographystyle{apsrev4-1}
\bibliography{reference}

\end{document}